\documentclass[12pt]{article}
\usepackage{amsfonts}
\usepackage{amsmath}

\begin{document}

\begin{titlepage}
\title{\bf Mechanics Systems on Para-K\"{a}hlerian Manifolds of Constant J-Sectional Curvature}
\author{Mehmet Tekkoyun \footnote{tekkoyun@pamukkale.edu.tr} \\
 {\small Department of Mathematics, Pamukkale University,}\\
{\small 20070 Denizli, Turkey}}
\date{\today}
\maketitle

\begin{abstract}

The goal of this paper is to present Euler-Lagrange and
Hamiltonian equations on  $\mathbf{R}_{n}^{2n}$ which is a model
of para-K\"{a}hlerian manifolds of constant J-sectional curvature.
In conclusion, some differential geometrical and physical results
on the related mechanic systems have been given.



\end{abstract}
\end{titlepage}

\section{Introduction}

It is well-known that Modern Differential Geometry has an important role to
obtain different types of Lagrangian and Hamiltonian formalisms of Classical
Mechanics. Moreover, it is possible to see a lot of studies in the suitable
fields. One may say that Lagrangian and Hamiltonian systems are
characterized by a convenient vector field $X$ defined on the tangent and
cotangent bundles which are phase-spaces of velocities and momentum of a
given configuration manifold. If $Q$ is an $m$-dimensional configuration
manifold and $L:TQ\rightarrow R$ is a regular Lagrangian function, then
there is a unique vector field $X$ on $TQ$ such that
\begin{equation}
i_{X_{L}}\omega _{L}=dE_{L},  \label{1.1}
\end{equation}%
where $\omega _{L}$ is the symplectic form and $E_{L}$ is energy associated
to $L$. The so-called Euler-Lagrange vector field $X$ is a semispray (or
\textit{second order differential equation}) on $Q$ since its integral
curves are the solutions of the Euler-Lagrange equations. The triple $%
(TQ,\omega _{L},L)$ is called \textit{Lagrangian system} on the tangent
bundle $TQ.$ If $H:T^{\ast }Q\rightarrow R$ is a regular Hamiltonian
function then there is a unique vector field $X_{H}$ on $T^{\ast }Q$ such
that
\begin{equation}
i_{X_{H}}\omega =dH  \label{1.2}
\end{equation}%
where $\omega $ is the symplectic form and $H$ stands for Hamiltonian
function. The paths of the so-called Hamiltonian vector field $X_{H}$ are
the solutions of the Hamiltonian equations. The triple $T^{\ast }Q,\omega
,H) $ is called \textit{Hamiltonian system} on the cotangent bundle $T^{\ast
}Q$ endowed with symplectic form $\omega $.

From the before some studies given in \cite{crampin, deleon,
crampin1, tekkoyun1, tekkoyun2, tekkoyun3, tekkoyun4} and there
in; we know that time-dependent or not, constraint, real, complex
and paracomplex analogues of the Lagrangian and Hamiltonian
systems have detailed been researced. But, we see that is not
mentioned about Lagrangian and Hamiltonian mechanics systems on
constant $J$-sectional curvature. Therefore, in this paper we
present the
Euler-Lagrange equations and Hamiltonian equations on a model of para-K\"{a}%
hlerian manifolds of constant $J$-sectional curvature and derive
differential geometrical and physical conclusions on related
dynamics systems.

In this paper, all the manifolds and geometric objects are
$C^{\infty }$ and
the Einstein summation convention is in use. Also, $\mathbf{R}$, $\mathcal{F}%
(M)$, $\chi (M)$ and $\Lambda ^{1}(M)$ denote the set of real numbers, the
set of functions on $M$, the set of vector fields on $M$ and the set of
1-forms on $M$, respectively.

\section{Para-K\"{a}hlerian Manifolds of Constant $J$-Sectional Curvature}

\textbf{Definition 1 }\cite{bejan, cruceanu}\textbf{\ : }Let a manifold $M$
be endowed with an almost product structure $J\neq \mp Id$; which is a (1;
1)-tensor field such that $J^{2}=Id$: We say that $(M,J)$ (resp.$(M,J,g)$)
is an almost product (resp. almost Hermitian) manifold, where $g$ is a
semi-Riemannian metric on $M$ with respect to which $J$ is skew-symmetric,
that is
\begin{equation}
g(JX,Y)+g(X,JY)=0,\forall X,Y\in \chi (M)  \label{2.1}
\end{equation}%
Then $(M,J,g)$ is para-K\"{a}hlerian if $J$ is parallel with respect to the
Levi-Civita connection.

Let $(M,J,g)$ be a para-K\"{a}hlerian manifold and let denote the curvature
(0, 4)-tensor field by
\begin{equation}
R(X,Y,Z,V)=g(R(X,Y)Z,V);\forall X,Y,Z,V\in \chi (M)  \label{2.2}
\end{equation}%
where the Riemannian curvature (1, 3)-tensor field associated to the
Levi-Civita connection $\nabla $ of $g$ is given by $R=$ [$\nabla ,\nabla $]
-$\nabla _{\left[ \text{ },~\text{\ }\right] }$.

Then

\begin{equation}
\begin{array}{c}
R(X,Y,Z,V)=-R(Y,X,Z,V)=-R(X,Y,V,Z)=R(JX,JY,Z,V) \\
\text{and}\underset{\sigma }{\text{ }\sum }R(X,Y,Z,V)=0,%
\end{array}
\label{2.3}
\end{equation}

where $\sigma $ denotes the sum over all cyclic permutations. We know that
the following (0,4)-tensor field is defined by%
\begin{equation}
R_{0}(X,Y,Z,V)=\frac{1}{4}\left\{
\begin{array}{c}
g(X,Z)g(Y,V)-g(X,V)g(Y,Z)-g(X,JZ)g(Y,JV) \\
+g(X,JV)g(Y,JZ)-2g(X,JY)g(Z,JV)%
\end{array}%
\right\} ,  \label{2.4}
\end{equation}

where $\forall X,Y,Z,V\in \chi (M).$ For any $p\in M$, a subspace $S\subset
T_{p}M$ is called non-degenerate if $g$ restricted to $S$ is non-degenerate.
If $\left\{ u,v\right\} $ is a basis of a plane $\sigma \subset T_{p}M$,
then $\sigma $ is

non-degenerate iff $g(u,u)g(v,v)-[g(u,v)]^{2}\neq 0$. In this case the
sectional curvature of $\sigma $= span$\left\{ u,v\right\} $ is

\begin{equation}
k(\sigma )=\frac{R(u,v,u,v)}{g(u,u)g(v,v)-[g(u,v)]^{2}}  \label{2.5}
\end{equation}

From (\ref{2.1}) it follows that $X$ and $JX$ are orthogonal for any $X\in $
$\chi (M)$. By a $J$-plane we mean a plane which is invariant by $J$. For
any $p\in M$, a vector $u$ $\in T_{p}M$ is isotropic provided $g(u,u)=0$. If
$u$ $\in T_{p}M$ is not isotropic, then the sectional curvature $H(u)$ of
the $J$-plane span$\left\{ u,Ju\right\} $ is called the $J$-sectional
curvature defined by $u.$ When $H(u)$ is constant,

then $(M,J,g)$ is called of constant $J$-sectional curvature, or a para-K%
\"{a}hlerian space form.

\textbf{Theorem 1:} Let $(M,J,g)$ be a para-K\"{a}hlerian manifold such that
for each $p\in M$, there exists $c_{p}\in R$ satisfying $H(u)=c_{p}$ for $u$
$\in T_{p}M$ \ such that $g(u,u)g(Ju,Ju)\neq 0.$Then the Riemann-
Christoffel tensor $R$ satisfies $R=cR_{0},$ where $c$ is the function
defined by $p\rightarrow c_{p}.$ And conversely.

\textbf{Definition 2: }A para-K\"{a}hlerian manifold $(M,J,g)$ is said to be
of constant paraholomorphic sectional curvature $c$ if it satisfies the
conditions of \textbf{Theorem 1}.

\textbf{Theorem 2:} Let $(M,J,g)$ be a para-K\"{a}hlerian manifold with $%
dimM>2$. Then the following properties are equivalent:

1) $M$ is a space of constant paraholomorphic sectional curvature $c=0$

2) The Riemann- Christoffel tensor curvature tensor $R$ has the expression%
\begin{equation}
R(X,Y,Z,V)=0,\forall X,Y,Z,V\in \chi (M).  \label{2.6}
\end{equation}

Let $(x_{i},y_{i})$ be a real coordinate system on a neighborhood $U$ of any
point $p$ of $\mathbf{R}_{n}^{2n},$ and $\{(\frac{\partial }{\partial x_{i}}%
)_{p},(\frac{\partial }{\partial y_{i}})_{p}\}$ and $%
\{(dx_{i})_{p},(dy_{i})_{p}\}$ natural bases over $\mathbf{R}$ of the
tangent space $T_{p}(\mathbf{R}_{n}^{2n})$ and the cotangent space $%
T_{p}^{\ast }(\mathbf{R}_{n}^{2n})$ of $\mathbf{R}_{n}^{2n},$ respectively.

The space $(\mathbf{R}_{n}^{2n},g,J),$ is the model of the para-K\"{a}%
hlerian space forms of dimension $2n\geq 2$ and paraholomorphic sectional
curvature $c=0,$ where $g$ is the metric

\begin{equation}
g=dx_{i}\otimes dy_{i}+dy_{i}\otimes dx_{i},  \label{2.7}
\end{equation}

and $J$ the almost product structure%
\begin{equation}
J=\frac{\partial }{\partial x_{i}}\otimes dx_{i}-\frac{\partial }{\partial
y_{i}}\otimes dx_{i}.  \label{2.8}
\end{equation}

Then we have
\begin{equation}
J(\frac{\partial }{\partial x_{i}})=\frac{\partial }{\partial x_{i}},\text{ }%
J(\frac{\partial }{\partial y_{i}})=-\frac{\partial }{\partial y_{i}}.
\label{2.9}
\end{equation}%
The dual endomorphism $J^{\ast }$ of the cotangent space $T_{p}^{\ast }(%
\mathbf{R}_{n}^{2n})$ at any point $p$ of manifold $\mathbf{R}_{n}^{2n}$
satisfies $J^{\ast 2}=Id$ and is defined by
\begin{equation}
J^{\ast }(dx_{i})=dx_{i},\text{ }J^{\ast }(dy_{i})=-dy_{i}.  \label{2.10}
\end{equation}

\section{Lagrangian Mechanics Systems}

Here, we introduce Euler-Lagrange equations on para-K\"{a}hlerian
manifolds of constant J-sectional curvature
$(\mathbf{R}_{n}^{2n},g,J)$.

Given by $J$ almost product structure and by $(x_{i},y_{i})$ the
coordinates of $\mathbf{R}_{n}^{2n}$. Let semispray be a vector
field as follows:
\begin{equation}
\xi =X_{i}\frac{\partial }{\partial x_{i}}+Y_{i}\frac{\partial }{\partial
y_{i}},\text{ }X_{i}=\overset{.}{x_{i}}=y_{i},\text{ }Y_{i}=\overset{.}{y_{i}%
}.  \label{3.1}
\end{equation}%
By \textit{Liouville vector field} on para-K\"{a}hlerian space form $(%
\mathbf{R}_{n}^{2n},g,J),$ we call the vector field determined by $V=J\xi $
and calculated by
\begin{equation}
J\xi =X_{i}\frac{\partial }{\partial x_{i}}-Y_{i}\frac{\partial }{\partial
y_{i}},  \label{3.2}
\end{equation}%
Denote $T$ by \textit{the kinetic energy} and $P$ by \textit{the
potential energy of mechanics system} on para-K\"{a}hlerian space
of constant $J$-sectional curvature,. Then we write by $L=T-P$
\textit{Lagrangian function }and by $E_{L}=V(L)-L$ \textit{the
energy function} associated $L$.

Operator $i_{J}$\ defined by\
\begin{equation}
i_{J}:\wedge ^{2}\mathbf{R}_{n}^{2n}\rightarrow \wedge ^{1}\mathbf{R}%
_{n}^{2n}  \label{3.3}
\end{equation}%
is called the interior product with $J$, or sometimes the insertion
operator, or contraction by $J.$ The exterior vertical derivation $d_{J}$ is
defined by
\begin{equation}
d_{J}=[i_{J},d]=i_{J}d-di_{J},  \label{3.4}
\end{equation}%
where $d$ is the usual exterior derivation. For almost product structure $J$
determined by (\ref{2.9}), the closed para-K\"{a}hlerian form is the closed
2-form given by $\Phi _{L}=-dd_{J}L$ such that
\begin{equation}
d_{J}=\frac{\partial }{\partial x_{i}}dx_{i}-\frac{\partial }{\partial y_{i}}%
dy_{i}:\mathcal{F}(\mathbf{R}_{n}^{2n})\rightarrow \wedge ^{1}\mathbf{R}%
_{n}^{2n}.  \label{3.5}
\end{equation}%
Thus we get
\begin{eqnarray}
\Phi _{L} &=&-\frac{\partial ^{2}L}{\partial x_{j}\partial x_{i}}%
dx_{j}\wedge dx_{i}-\frac{\partial ^{2}L}{\partial y_{j}\partial x_{i}}%
dy_{j}\wedge dx_{i}  \notag \\
&&+\frac{\partial ^{2}L}{\partial x_{j}\partial y_{i}}dx_{j}\wedge dy_{i}+%
\frac{\partial ^{2}L}{\partial y_{j}\partial y_{i}}dy_{j}\wedge dy_{i}.
\label{3.6}
\end{eqnarray}%
Then%
\begin{equation}
\begin{array}{c}
i_{\xi }\Phi _{L}=-X_{i}\frac{\partial ^{2}L}{\partial x_{j}\partial x_{i}}%
\delta _{i}^{j}dx_{i}+X_{i}\frac{\partial ^{2}L}{\partial x_{j}\partial x_{i}%
}dx_{j}-Y_{i}\frac{\partial ^{2}L}{\partial y_{j}\partial x_{i}}\delta
_{i}^{j}dx_{i}+X_{i}\frac{\partial ^{2}L}{\partial y_{j}\partial x_{i}}dy_{j}
\\
+X_{i}\frac{\partial ^{2}L}{\partial x_{j}\partial y_{i}}\delta
_{i}^{j}dy_{i}-Y_{i}\frac{\partial ^{2}L}{\partial x_{j}\partial y_{i}}%
dx_{j}+Y_{i}\frac{\partial ^{2}L}{\partial y_{j}\partial y_{i}}\delta
_{i}^{j}dy_{i}-Y_{i}\frac{\partial ^{2}L}{\partial y_{j}\partial y_{i}}%
dy_{j}.%
\end{array}
\label{3.7}
\end{equation}%
Because of the closed para-K\"{a}hlerian form $\Phi _{L}$ on
para-K\"{a}hlerian space form $(\mathbf{R}_{n}^{2n},g,J)$ is
para-symplectic structure, one may obtain
\begin{equation}
E_{L}=X_{i}\frac{\partial L}{\partial x_{i}}-Y_{i}\frac{\partial L}{\partial
y_{i}}-L,  \label{3.8}
\end{equation}%
and thus
\begin{equation}
\begin{array}{ll}
dE_{L}= & X_{i}\frac{\partial ^{2}L}{\partial x_{j}\partial x_{i}}%
dx_{j}-Y_{i}\frac{\partial ^{2}L}{\partial x_{j}\partial y_{i}}dx_{j}-\frac{%
\partial L}{\partial xj}dx_{j} \\
& +X_{i}\frac{\partial ^{2}L}{\partial y_{j}\partial x_{i}}dy_{j}-Y_{i}\frac{%
\partial ^{2}L}{\partial y_{j}\partial y_{i}}dy_{j}-\frac{\partial L}{%
\partial yj}dy_{j}%
\end{array}
\label{3.9}
\end{equation}%
Taking care of $i_{\xi }\Phi _{L}=dE_{L}$, we have
\begin{equation}
\begin{array}{l}
-X_{i}\frac{\partial ^{2}L}{\partial x_{j}\partial x_{i}}dx_{j}-Y_{i}\frac{%
\partial ^{2}L}{\partial y_{j}\partial x_{i}}dx_{j}+\frac{\partial L}{%
\partial xj}dx_{j} \\
+X_{i}\frac{\partial ^{2}L}{\partial x_{j}\partial y_{i}}dy_{j}+Y_{i}\frac{%
\partial ^{2}L}{\partial y_{j}\partial y_{i}}dy_{j}+\frac{\partial L}{%
\partial yj}dy_{j}=0.%
\end{array}
\label{3.10}
\end{equation}%
If the curve $\alpha :\mathbf{I\subset R}\rightarrow \mathbf{R}_{n}^{2n}$ be
integral curve of $\xi ,$ which satisfies
\begin{equation}
\begin{array}{l}
-[X_{i}\frac{\partial ^{2}L}{\partial x_{j}\partial x_{i}}+Y_{i}\frac{%
\partial ^{2}L}{\partial y_{j}\partial x_{i}}]dx_{j}+\frac{\partial L}{%
\partial xj}dx_{j} \\
+[X_{i}\frac{\partial ^{2}L}{\partial x_{j}\partial y_{i}}+Y_{i}\frac{%
\partial ^{2}L}{\partial y_{j}\partial y_{i}}]dy_{j}+\frac{\partial L}{%
\partial yj}dy_{j}=0.%
\end{array}
\label{3.11}
\end{equation}%
it follows equations

\begin{equation}
\frac{\partial }{\partial t}\left( \frac{\partial L}{\partial x_{j}}\right) -%
\frac{\partial L}{\partial x_{j}}=0,\ \frac{\partial }{\partial t}\left(
\frac{\partial L}{\partial y_{j}}\right) +\frac{\partial L}{\partial y_{j}}=0
\label{3.12}
\end{equation}%
so-called \textit{Euler-Lagrange equations }whose solutions are the paths of
the semispray $\xi $ on para-K\"{a}hlerian space form $(\mathbf{R}%
_{n}^{2n},g,J)$. Finally one may say that the triple $(\mathbf{R}%
_{n}^{2n},\Phi _{L},\xi )$ is\textit{\ mechanical system} on para-K\"{a}%
hlerian manifolds of constant J-sectional curvature
$(\mathbf{R}_{n}^{2n},g,J).$Therefore we say

\textbf{Proposition 1: }Let $J$ almost product structure on
para-K\"{a}hlerian space of constant $J$-sectional curvature
$(\mathbf{R}_{n}^{2n},g,J).$Also let $(f_{1},f_{2})$ be natural
bases of $\mathbf{R}_{n}^{2n}.$Then it follows

\begin{equation*}
\begin{array}{cc}
J(f_{1})-f_{1}=0\Longleftrightarrow & \overset{.}{f}_{1,L}-f_{1,L}=0, \\
J(f_{2})+f_{2}=0\Longleftrightarrow & \overset{.}{f}_{2,L}+f_{2,L}=0,%
\end{array}%
\end{equation*}

where $f_{1,L}=\frac{\partial L}{\partial x_{i}},$ $\ f_{2,L}=\frac{\partial
L}{\partial y_{i}},$ $\overset{.}{f}_{1,L}=\frac{\partial }{\partial t}(%
\frac{\partial L}{\partial x_{i}}),$ $\ \overset{.}{f}_{2,L}=\frac{\partial
}{\partial t}(\frac{\partial L}{\partial y_{i}}).$

\section{Hamiltonian Mechanics Systems}

Here, we present Hamiltonian equations on para-K\"{a}hlerian
manifolds of constant $J$-sectional curvature
$(\mathbf{R}_{n}^{2n},g,J)$.

Let $J^{\ast }$ be an almost product structure defined by (\ref{2.10}) and $%
\lambda $ Liouville form determined by $J^{\ast }(\omega )=\frac{1}{2}%
y_{i}dx_{i}-\frac{1}{2}x_{i}dy_{i}$ such that $\omega =\frac{1}{2}%
y_{i}dx_{i}+\frac{1}{2}x_{i}dy_{i}$ 1-form on $\mathbf{R}_{n}^{2n}.$ If $%
\Phi =-d\lambda $ is closed para-K\"{a}hlerian form$,$ then it is also a
para-symplectic structure on $\mathbf{R}_{n}^{2n}$.

Let $(\mathbf{R}_{n}^{2n},g,J)$\textbf{\ }be para-K\"{a}hlerian
manifolds of constant J-sectional curvature with closed
para-K\"{a}hlerian form $\Phi $. Suppose that Hamiltonian vector
field $Z_{H}$ associated to Hamiltonian energy $H$ is given by
\begin{equation}
Z_{H}=X_{i}\frac{\partial }{\partial x_{i}}+Y_{i}\frac{\partial }{\partial
y_{i}}.  \label{4.1}
\end{equation}

For the closed para-K\"{a}hlerian form $\Phi $ on $\mathbf{R}_{n}^{2n},$ we
have
\begin{equation}
\Phi =-d\lambda =-d(\frac{1}{2}y_{i}dx_{i}-\frac{1}{2}x_{i}dy_{i})=dx_{i}%
\wedge dy_{i}.  \label{4.2}
\end{equation}%
Then it follows
\begin{equation}
i_{Z_{H}}\Phi =\Phi (Z_{H})=-Y_{i}dx_{i}+X_{i}dy_{i}.  \label{4.3}
\end{equation}%
Otherwise, one may calculate the differential of Hamiltonian
energy as follows:
\begin{equation}
dH=\frac{\partial H}{\partial x_{i}}dx_{i}+\frac{\partial H}{\partial y_{i}}%
dy_{i}.  \label{4.4}
\end{equation}%
From (\ref{4.3}) and (\ref{4.4}) with respect to $i_{Z_{H}}\Phi
=dH,$ we find para-Hamiltonian vector field on para-K\"{a}hlerian
space of constant $J$-sectional curvature to be
\begin{equation}
Z_{H}=\frac{\partial H}{\partial y_{i}}\frac{\partial }{\partial x_{i}}-%
\frac{\partial H}{\partial x_{i}}\frac{\partial }{\partial y_{i}}.
\label{4.5}
\end{equation}

Suppose that the curve
\begin{equation}
\alpha :I\subset \mathbf{R}\rightarrow \mathbf{R}_{n}^{2n}  \label{4.6}
\end{equation}%
be an integral curve of Hamiltonian vector field $Z_{H},$ i.e.,
\begin{equation}
Z_{H}(\alpha (t))=\overset{.}{\alpha },\,\,t\in I.  \label{4.7}
\end{equation}%
In the local coordinates we have
\begin{equation}
\alpha (t)=(x_{i}(t),y_{i}(t)),  \label{4.8}
\end{equation}%
and
\begin{equation}
\overset{.}{\alpha }(t)=\frac{dx_{i}}{dt}\frac{\partial }{\partial x_{i}}+%
\frac{dy_{i}}{dt}\frac{\partial }{\partial y_{i}}.  \label{4.9}
\end{equation}%
Now, by means of (\ref{4.7}), from (\ref{4.5}) and (\ref{4.9}), we deduce
the equations so-called \textit{para}-\textit{Hamiltonian equations}
\begin{equation}
\frac{dx_{i}}{dt}=\frac{\partial H}{\partial y_{i}},\frac{dy_{i}}{dt}=-\frac{%
\partial H}{\partial x_{i}}.  \label{4.10}
\end{equation}%
In the end, we may say to be \textit{para-mechanical system }$(\mathbf{R}%
_{n}^{2n},\Phi ,Z_{H})$ triple on para-K\"{a}hlerian manifolds of constant J-sectional curvature $(\mathbf{R}%
_{n}^{2n},g,J).$

\section{Conclusion}

From the study,\textbf{\ }we obtain that Lagrangian and
Hamiltonian formalisms in generalized Classical Mechanics and
field theory can be intrinsically characterized on
$(\mathbf{R}_{n}^{2n},g,J)$ being a model of
para-K\"{a}hlerian space of constant $J$-sectional curvature$.$ So, the paths of semispray $\xi $ on $%
\mathbf{R}_{n}^{2n}$ are the solutions of the Euler-Lagrange
equations given by (\ref{3.12}) on the mechanical system
$(\mathbf{R}_{n}^{2n},\Phi _{L},\xi
)$. Also, the solutions of the Hamiltonian equations determined by (\ref%
{4.10}) on the mechanical system $(\mathbf{R}_{n}^{2n},\Phi ,Z_{H})$ are the
paths of vector field $Z_{H}$ on $\mathbf{R}_{n}^{2n}$.

\end{document}